\documentclass{article}
\usepackage{spconf,amsmath,graphicx}
\usepackage{cite}
\usepackage{spconf,amsmath,graphicx}
\usepackage{amssymb}
\usepackage{mathrsfs}
\usepackage{amsmath}
\usepackage{comment} 
\usepackage{here}
\usepackage{bm}
\usepackage{url}
\usepackage{booktabs}
\usepackage{multirow}
\usepackage{enumitem}

\title{Echo-aware Adaptation of Sound Event Localization and Detection \\in Unknown Environments}

\vspace{-12pt}
\name{
Masahiro Yasuda$^{\dagger}$,
Yasunori Ohishi$^{\dagger}$,
Shoichiro Saito$^{\dagger}$
}
\vspace{-12pt}
\address{$^\dagger$NTT Corporation, Japan
\vspace{-12pt}
}

\begin{document}
\ninept

\maketitle

\begin{abstract}
Our goal is to develop a sound event localization and detection (SELD) system that works robustly in unknown environments. 
A SELD system trained on known environment data is degraded in an unknown environment due to environmental effects such as reverberation and noise not contained in the training data. 
Previous studies on related tasks have shown that domain adaptation methods are effective when data on the environment in which the system will be used is available even without labels. However adaptation to unknown environments remains a difficult task.
In this study, we propose echo-aware feature refinement (EAR) for SELD, which suppresses environmental effects at the feature level by using additional spatial cues of the unknown environment obtained through measuring acoustic echoes.
FOA-MEIR
\footnote{\scriptsize{The dataset will be available by the date of ICASSP2022 at \url{https://github.com/nttrd-mdlab/seld-foa-meir}}}
, an impulse response dataset containing over 100 environments, was recorded to validate the proposed method. Experiments on FOA-MEIR show that the EAR effectively improves SELD performance in unknown environments.

\end{abstract}

\begin{keywords}
sound event localization and detection (SELD), model adaptation, multi-modal, deep neural network (DNN).
\end{keywords}

\vspace{-5pt}
\section{Introduction}
\vspace{-3pt}
\label{sec:intro}
In our surrounding environment, there are various sounds such as speech, machine activity, and animal sounds. 
A system capable of detecting such sounds will provides us various valuable applications such as automatic driving to detect unseen dangers~\cite{car,SmartCar1,SmartCar2}, detection of crimes in the dark~\cite{surveillance,drone}, and support for the safety of pedestrians.
A key technique for such applications is sound event localization and detection (SELD),which combines direction of arrival (DOA) estimation and sound event detection (SED).

In a recent competition of SELD, many algorithms using a deep neural networks (DNNs) as a regression function of DOA and classification function of SED have achieved high performance~\cite{dcase2019sota,dcase2020sota,dcase2021sota}. To train such DNNs, a large amount of annotated data is needed that contains various sound events occurring in various directions around the microphone. 
In the SELD problem settings considered to date, the system is trained using the SELD dataset recorded in up to 11 environments, and then performance is evaluated using data recorded in the same environments.

On the other hand, since users will utilize the SELD system in any environment they want in real-world applications, the system should work robustly in environments not included in the training data, referred to as unknown environments in this paper.
The simplest way to train a robust SELD system for any environment is to record complete IR datasets in many environments and use them for training, as used in conventional SELD methods. However, recording and annotating three-dimensional sound data incurs a huge cost. In particular, since countless factors affect the directional information of sound, such as the arrangement of objects in a room, distance from the walls to the microphone, and building materials, covering all of these combinations is unrealistic.

Several previous studies on SED and DOA estimation have addressed this issue by using domain adaptation approaches~\cite{doaadap1,doaadap2,sedadap,sedadap2,sedadap3}. 
In SED, it is shown that domain adversarial training (DAT) is effective in keeping the performance at the target domain~\cite{sedadap}. On the other hand, in DOA estimation, it has been reported that DAT does not work effectively, and weak labels on DOA are required for adaptation to target domain~\cite{doaadap2}. These domain adaptation methods in SED and DOA estimation require the data recorded in the target domain, so they have difficulty adapting models to unknown environments.

Another strategy for adapting to unknown environments is utilizing sounds that the system ``hears'' during inference. For robot audition, noise-robust sound event detection has been proposed utilizing observed background noise.~\cite{adapnoise1}. 
In addition, for SELD, not only noise but also reverberation must be helpful cues. 
In particular, the reflections of the sound emitted by the system itself, which we call `echo' in this paper, has been reported to provide a wealth of spatial cues such as room's shape and the arrangement of reflectors~\cite{echo1,echo2,echo3}. 
Combining it with the SELD system is a promising strategy for adaption to unknown environments.

To take advantage of the spatial cues provided by echoes, we propose an echo-aware feature refinement (EAR) for SELD.
The proposed system incorporates a feature refinement mechanism conditioned on embeddings extracted from echoes to suppress environmental effects that cause performance degradation in an unknown environment.
This refinement mechanism is trained using the DAT framework so that the refinement features are indistinguishable from anechoic ones.
For our new task, we also recorded a new dataset: multi-environment impulse response recordings with a first-order ambisonic microphone (FOA-MEIR). This dataset combines comprehensive IR recordings in an anechoic room and sparse IR recordings in nearly 100 real environments, which exceed the total number of environments in the SELD dataset published so far~\cite{dataset2019,dataset2020,dataset2021}. Experimental results on the FOA-MEIR dataset show that the EAR effectively improves the performance of SELD in unknown environments and outperforms the baseline method without EAR.
\vspace{-9pt}
\section{Related studies}
\label{sec:relatedstudies}
\vspace{-5pt}
\subsection{Overview of SELD}
\vspace{-4pt}
SELD 
task has been attracting particular attention since it was taken up as a challenge task in DCASE 2019~\cite{dataset2019}.
Recently, more real-world like problem settings of the SELD are attempted~\cite{seldproblem}. In the task of DCASE 2019, SELD for stationary polyphonic sound sources recorded in five environments was handled, and it was expanded to the moving sound sources in 2020, and the unknown directional interferers were added in 2021~\cite{dataset2019,dataset2020,dataset2021}. However, in any problem setting, training data and test data are generated using IR recorded in the same environment, and the adaptation to an unknown environment has not been examined yet. 


\vspace{-5pt}
\subsection{Domain adaptation in related tasks}
\vspace{-2pt}
In SED and acoustic scene classification, several methods based on domain adaptive learning have been proposed to bridge the gap between the training data and the target domain~\cite{sedadap,sedadap2,sedadap3}.
In the domain adaptation utilizing DAT, a discriminator is introduced to distinguish the features obtained from the source and target domains.  In contrast, the feature extractor is trained adversarially to trick the discriminator, resulting in domain-invariant features are extracted.
In DOA estimation, a method was proposed to adapt a DNN trained with labeled data recorded in an anechoic room to unlabeled data recorded in a reverberant room was proposed~\cite{doaadap1}. This method estimates DOA as a classification problem, and performs domain adaptation by retraining the DNN model so as to minimize the entropy of output. While entropy minimization-based methods are limited to DOA estimation of a single source, He {\it et al.} proposed a domain adaptation method that can apply to DOA estimation of multiple sources~\cite{doaadap2}. 

\vspace{-5pt}
\section{Proposed method}
\vspace{-3pt}
Our goal is to develop a SELD system that works robustly even in unknown environments, only using training data and its labels of known environments. This section describes the proposed method to achieve such a system.

Here, we define some notations.
Define $\mathcal{E}=\{e_1,\ldots,e_N\}$ as the $N$ known environments. The unknown environment $\mathcal{E}^*\not\subset\mathcal{E}$ is defined as any environment that is not included in $\mathcal{E}$. The observed signals in the known and unknown environments are denoted as $\bm{x}_{\mathcal{E}}$ and $\bm{x}_{\mathcal{E}^*}$, respectively.
Let SELD prediction function represent as $\mathcal{M}$ that predict SELD ground truth labels $Y_{\mathcal{E}}=\{\bm{y}_1,\ldots,\bm{y}_K\}$ from corresponding training data $X_{\mathcal{E}}=\{\bm{x}_1,\ldots,\bm{x}_K\}$. Here, $\bm{y}$ include both SED and DOA label as $\bm{y}=\{y_{\mbox{\scriptsize{SED}}},y_{\mbox{\scriptsize{DOA}}}\}$. In addition, as a basic architecture~\cite{dataset2020}, $\mathcal{M}$ consists of the feature extractor $\mathcal{F}$, SED classifier $\mathcal{C}$ and DOA regression function $\mathcal{D}$.

\vspace{-5pt}
\subsection{Basic concept}
\vspace{-4pt}
\label{sec:basic}
The performance of a SELD system trained for particular known environment will degrade in an unknown environment.
This problem is known as domain shift.
A feature extractor trained for known environments does not work properly in an unknown environment due to environmental effects such as noise and reverberation, resulting in a shift of feature statistics and consequent performance degradation.

\begin{figure}[t!]
\begin{center}
  \includegraphics[width=\linewidth]{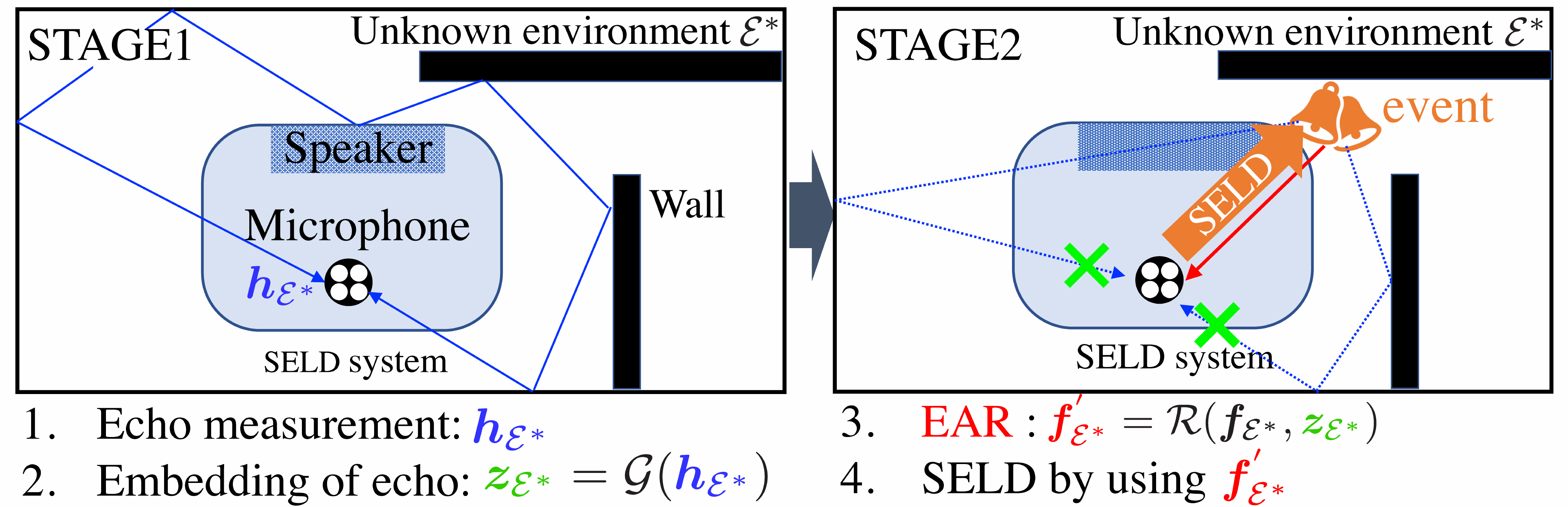}
 \vspace{-18pt}
  \caption{Basic concept of echo-aware feature refinement (EAR).}
  \label{fig:ear}
  \end{center}
  \vspace{-25pt}
 \end{figure}

To address this problem, we propose echo-aware feature refinement (EAR).
Fig.~\ref{fig:ear} shows the two-stage inference procedure.
The first stage is the echo measurement in an unknown environment.
The sound source for the echo measurement is assumed to be the system's own sound, such as the startup sound.
The observed echo $\bm{h}_{\mathcal{E}^{*}}$ is embedded into $\bm{z}_{\mathcal{E}^{*}}$ via the encoder $\mathcal{G}$.
Since echoes hold a wealth of spatial cues about the surrounding environment~\cite{echo1,echo2,echo3}, such information about the unknown environment will be embedded in $\bm{z}_{\mathcal{E}^{*}}$.

The second stage is SELD.
To begin with, the feature extractor $\mathcal{F}$ extracts the feature $\bm{f}_{\mathcal{E}^
{*}}$ from the observed signal $\bm{x}_{\mathcal{E}^{*}}$ containing sound events occurring in the surroundings.
To suppress environmental effects such as noise and reverberation from $\bm{f}_{\mathcal{E}^{*}}$, we utilize the spatial cues embedded in $\bm{z}_{\mathcal{E}^{*}}$.
The refinement of $\bm{f}_{\mathcal{E}^{*}}$ utilizing $\bm{z}_{\mathcal{E}^{*}}$, that is EAR, is performed by the following equation:
\vspace{-1pt}
\begin{equation}
\label{eq:refinement}\bm{f}_{\mathcal{E}^{*}}^{'} = \mathcal{R}(\bm{f}_{\mathcal{E}^{*}}, \bm{z}_{\mathcal{E}^{*}}).
\vspace{-1pt}
\end{equation}
Here, $\mathcal{R}$ is the refinement function. Finally, on the basis of the obtained $\bm{f}^{'}_{\mathcal{E}^{*}}$, SELD is performed by the SED classifier $\mathcal{C}$ and the DOA regression function $\mathcal{D}$.

The above two-stage inference procedure requires training the feature extractor $\mathcal{F}$, encoder $\mathcal{G}$, and refinement function $\mathcal{R}$ using only known environment data.
For this training, we first prepare the paired data consisting of observed echo $\bm{h}_{\mathcal{E}}$ and sound event observation signals $\bm{x}_{\mathcal{E}}$ recorded in multiple known environments.  In addition, we collect paired data ($\bm{h}_{\mathcal{E}}$,$\bm{x}_{\mathcal{E}}$) in an anechoic room. 

To train the system by using this training data, we adopt the DAT framework. 
Note that, since the data in unknown environments $\mathcal{E}^{*}$, i.e., the true target domain, is not available, we set the anechoic room as the source domain and the other environment as the target domain. In training, paired data ($\bm{h}_{\mathcal{E}}$,$\bm{x}_{\mathcal{E}}$) is used to obtain the refined feature $\bm{f}^{'}_{\mathcal{E}}$ on the basis of Eq. (\ref{eq:refinement}). The domain classifier $\mathcal{H}$, which is used only during training, takes $\bm{f}^{'}_{\mathcal{E}}$ as input and classifies the domain, reverberant or anechoic.
By training $\mathcal{F}$,$\mathcal{G}$,$\mathcal{R}$ adversarially to degrade the performance of domain classification by $\mathcal{H}$, an environment-invariant feature $\bm{f}^{'}_{\mathcal{E}}$ can be obtained. 
Especially, the correction function $\mathcal{R}$ conditioned on $\bm{z}_{\mathcal{E}}$ is expected to acquire the ability to suppress the environmental effect at the feature level, on the basis of the spatial cues in surrounding environment embedded in $\bm{z}_{\mathcal{E}}$ (e.g., the arrangement of objects).

\vspace{-5pt}
\subsection{Data collection scheme for EAR}
\vspace{-3pt}
\label{sec:casm}
For the training of EAR, it is necessary to collect paired data of echo $\bm{h}_{\mathcal{E}}$ and observed signal containing sound event $\bm{x}_{\mathcal{E}}$ over many environments. Conventional datasets use comprehensively collected IR recordings at 5 to 11 environments to generate training data, but simply extending this in a larger number of environments is very costly. Especially when considering the development of real SELD devices, collecting a huge amount of data each time is not practical, as training data is required for each device.

To overcome this, we propose a data collection scheme, combination of comprehensive anechoic data and sparse multi-environment data (CASM), which is suitable for collecting multi-environment data.
Comprehensive anechoic data is the recordings of acoustic events or IRs in an anechoic room, including a comprehensive angle and distance of the sound source to the microphone. 
The sparse multi environment data, which are sound events or IRs recorded at a few variations of angles and distances over many environments, are expected to be useful in understanding how the environment affects the acoustic signal.

\vspace{-5pt}
\subsection{Implementation details}
\vspace{-3pt}
Fig.~\ref{fig:architecture} shows the network architecture of the proposed method.
Our system can be broadly divided into three parts; the main branch for the SELD task, the echo auto encoder (AE) to extract the embedding $z_{\mathcal{E}}$, and the domain classifier that is used only during training for DAT.
We adopt the baseline model of DCASE2020~\cite{dataset2020} as the main branch in order to validate the effect of EAR in combination with the standard SELD system.
The input features of the main branch are 4-channel logmel-spectrograms and 3-channel intensity vectors~\cite{intensityvector}. The feature extractor consisted of a multi-layer convolutional neural network (CNN)-block extracts the feature $\bm{f}_{\mathcal{E}}$.
$\bm{f}_{\mathcal{E}}$ is concatenated with $z_{\mathcal{E}}$ and inputted to the feature refinement function $\mathcal{R}$. Bi-directional gated recurrent unit (Bi-GRU) ~\cite{bigru} based architecture of $\mathcal{R}$ enables to be refined $\bm{f}_{\mathcal{E}}$ considering long-term dependencies of the observed signal such as the effect from the rear part reverberation.
The two branches for SED and DOA estimation take $\bm{f}^{'}_{\mathcal{E}}$ as input and output a prediction of SELD.
By using this output, the loss function for the SELD task is calculated using binary cross entropy (BCE) and mean square error (MSE) as follows:

\begin{figure}[t!]
\begin{center}
  \includegraphics[width=0.95\linewidth]{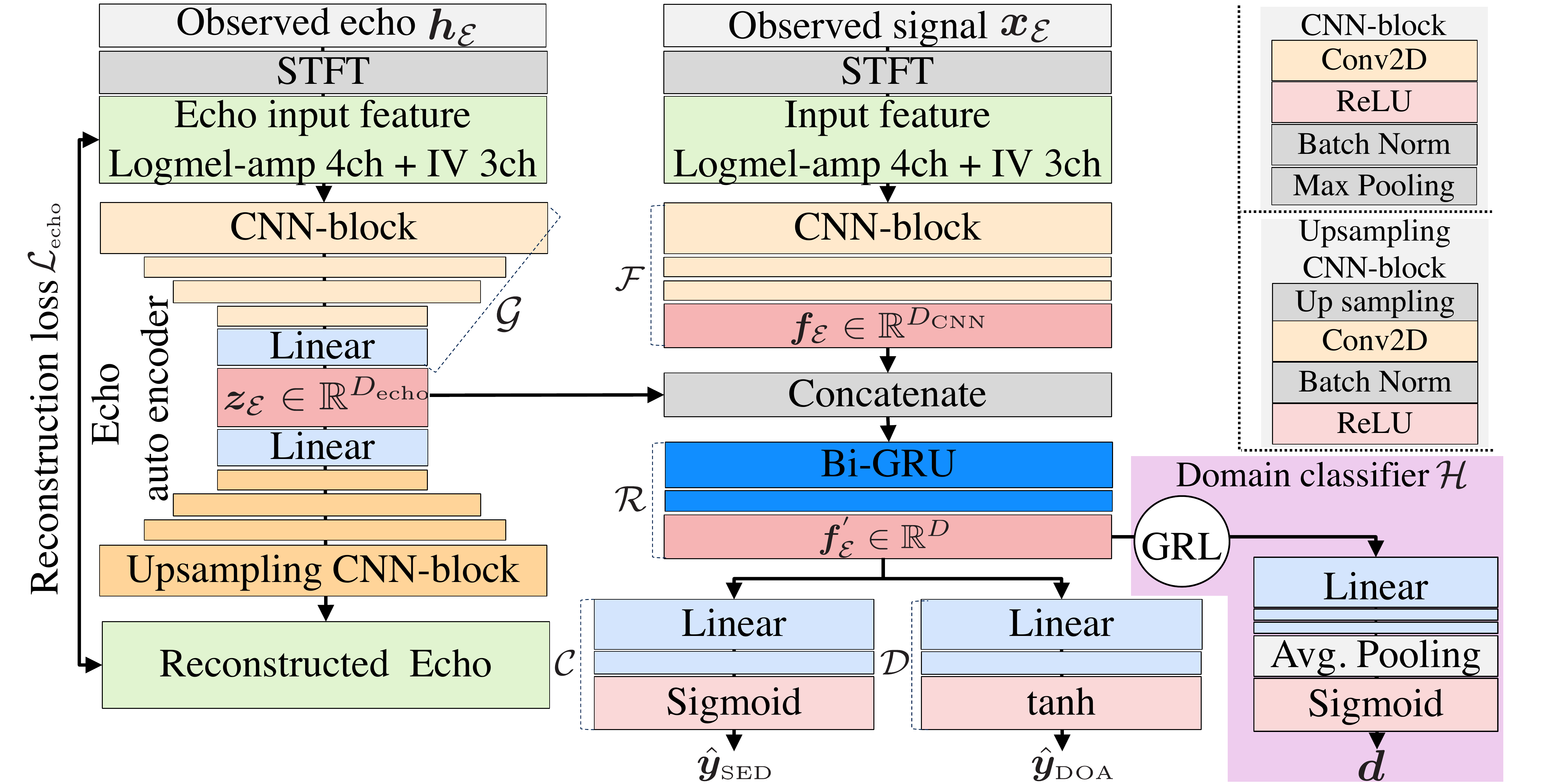}
 \vspace{-12pt}
  \caption{Network architecture of proposed method.}
  \label{fig:architecture}
  \end{center}
  \vspace{-25pt}
 \end{figure}
 
 \begin{figure*}[h!]
  \begin{center}
  \vspace{-6pt}
  \includegraphics[width=0.97\linewidth]{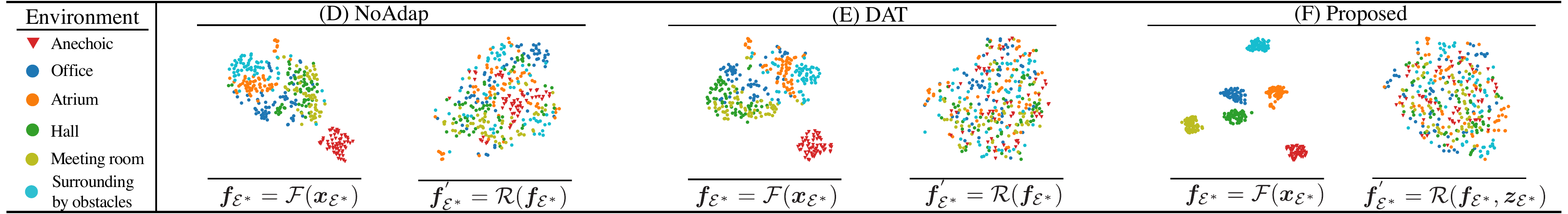}
  \vspace{-11pt}
  \caption{Comparison of t-SNE visualization of intermediate features.}
\vspace{-22pt}
\label{fig:embeddings}
  \end{center}
 \end{figure*}

\vspace{-12pt}
\begin{equation}
    \label{eq:lossseld}
    \mathcal{L}_{\mbox{\scriptsize{seld}}} = \mathcal{L}_{\mbox{\scriptsize{BCE}}}(\hat{\bm{y}}_{\mbox{\scriptsize{sed}}},\bm{y}_{\mbox{\scriptsize{sed}}}) + \lambda_{\mbox{\scriptsize{doa}}} \mathcal{L}_{\mbox{\scriptsize{MSE}}}(\hat{\bm{y}}_{\mbox{\scriptsize{doa}}},\bm{y}_{\mbox{\scriptsize{doa}}}) ,
\end{equation}
where $\lambda_{\mbox{\scriptsize{DOA}}}$ is the fixed balance parameter.

The domain classifier consists of a gradient reversal layer (GRL)~\cite{grl} and linear layers with a sigmoid activation function. It outputs a one-dimensional domain prediction $\hat{\bm{d}}$. Here, $\bm{d}=0,\,1$ denotes the anechoic and reverberant domains, respectively.
The GRL is introduced for DAT, which works as an identity layer in forward propagation and reverses the gradient in backpropagation with scale $\lambda_{\mbox{\scriptsize grl}}$.
We adopted BCE as the loss function $\mathcal{L}_{\mbox{\scriptsize domain}}$ to train the domain classifier.

The input of the echo AE is the observed echo $\bm{h}_{\mathcal{E}}$. 
Unlike the input signal $\bm{x}_{\mathcal{E}}$ for the main branch, $\bm{h}_{\mathcal{E}}$ is measured only once in each environment. 
The input feature of the echo AE, denoted as $Z_{\mathcal{E}}$, is extracted from $\bm{h}_{\mathcal{E}}$ with the same way as the input feature of the main branch.
The encoder part of the echo AE, which consists of multi-layer CNN-blocks and a linear layer, extracts $\bm{z}_{\mathcal{E}}$ from the input $Z_{\mathcal{E}}$.
The undesirable saddle point of DAT using GRL is that $\bm{z}_{\mathcal{E}}$ is trained as an environment-invariant embedding, and useful spatial cues for EAR are lost.
To avoid this, we set a reconstruction constraint on $\bm{z}_{\mathcal{E}}$.
For this constraint, we add the echo reconstruction loss $\mathcal{L}_{\mbox {\scriptsize echo}}$ that is MSE between the input feature $Z_{\mathcal{E}}$ and the reconstructed input feature $\tilde{Z}_{\mathcal{E}}$, to the training loss function.

The entire network is trained in an end-to-end manner on the basis of the following loss function:
\vspace{-3pt}
\begin{equation}
    \label{eq:loss}
    \mathcal{L} = \lambda_{\mbox{\scriptsize seld}}\mathcal{L}_{\mbox{\scriptsize seld}}+\lambda_{\mbox{\scriptsize domain}}\mathcal{L}_{\mbox{\scriptsize domain}}+\lambda_{\mbox{\scriptsize echo}}\mathcal{L}_{\mbox{\scriptsize echo}},
\end{equation}
\vspace{-5pt}
where $\lambda_{\mbox{\scriptsize seld}},\lambda_{\mbox{\scriptsize domain}}$ and  $\lambda_{\mbox{\scriptsize echo}}$ are fixed balance parameters.

\section{Experiments}
\vspace{-2pt}
\subsection{Dataset: FOA-MEIR}
\vspace{-2pt}
To set up a new problem of unknown environment adaptation of SELD, we collected a ``FOA-MEIR'' data set that records IR of many environments using a first-order ambisonic (FOA) microphone on the basis of the CASM scheme described in Sec.~\ref{sec:casm}. The data set consists of five subsets of IR recordings shown in Table \ref{tb:recording} and dry source recordings for SELD tasks.

The first subset, ``Anechoic'', contains 216 IR recordings in an anechoic room. 
These 216 IR recordings consist of a comprehensive combination of relative angles and distances between the sound source and the microphone.
The second subset, ``Reverb-S'', consists of sparse IR recordings in multiple reverberant environments.
It includes 96 environments such as offices, meeting rooms, halls, etc.
Each environment is different in at least one way: the room, the position of the microphone, and the arrangement of surrounding objects.
At each environment, three recording positions of IR were randomly selected from 216 angle and distance combinations that are the same as ``Anechoic''.
The third subset, ``Test'', contains 216 IR recordings in 5 unknown environments. 
The fourth subset, ``Echo'', consists of IR recordings recorded at 101 places where ``Reverb-S'' and ``Test'' were recorded, at the azimuth and elevation equals to 0 degree, and the distance of sound source from microphone equals to 150 cm.  This subset is used to simulate the observed echo $\bm{h}_{\mathcal{E}}$ described in Sec.~\ref{sec:basic}.
The fifth subset, ``Reverb-C'', is conventional like comprehensive reverberant IR recordings, which contains 216 IR recordings in two reverberant environments.
At each position of ``Reverb-S'',``Test'' and ``Reverb-C'', ambient noise was also recorded using the same FOA microphone that was used for IR recording. Ambient noise includes air conditioning, walking, talking, and so on. To synthesize a dataset for SELD using the above IR, dry sounds were recorded in an anechoic room using a monaural microphone. These dry sounds contain 12 different sound event classes, and each class has 20 variations of sound.

\begin{table}[t!]
\centering
\vspace{-10pt}

\caption{Specification of FOA-MEIR dataset}
\label{tb:recording}
\scalebox{0.78}[0.78]{
\begin{tabular}{@{}l|cccc|c@{}}
\toprule
Subset& Anechoic  & Reverb-S & Test &Echo& Reverb-C\\ \midrule
\# of environment & 1  & 96 & 5& 102&2\\
\# of IR / environment & 216  & 3 & 216 & 1&216 \\ \midrule
Azimuth range& $[-\pi,\pi)$&$[-\pi,\pi)$&$[-\pi,\pi)$ &$0$&$[-\pi,\pi)$\\
Azimuth interval&$10^{\circ}$&random&$10^{\circ}$&-&$10^{\circ}$\\ 
Elevation range&$[-\frac{\pi}{2},\frac{\pi}{2})$&$[-\frac{\pi}{2},\frac{\pi}{2})$&$[-\frac{\pi}{2},\frac{\pi}{2})$ &$0$&$[-\frac{\pi}{2},\frac{\pi}{2})$ \\
Elevation interval&$20^{\circ}$&random&$20^{\circ}$&-&$20^{\circ}$\\
Distance [cm] & 75,\,150 & 75,\,150 & 75,\,150 & 150&75,\,150\\ \midrule
Noise / environment &-&2.5 min&15 min&- & 15 min\\ 
\bottomrule
\end{tabular}
}
\vspace{-14pt}
\end{table}

\vspace{-15pt}
\begin{table}[t!]
\centering
\caption{Split settings of synthesis data for training and evaluation.}
\label{tb:split}
\scalebox{0.9}[0.9]{
\begin{tabular}{@{}l|cccc@{}}
\toprule
Split Name& IR & SNR & Length&\# of clips\\ \midrule
Train-rev&``Reverb-S'' &6 to 30dB&20 sec.&1920 \\ 
Train-anec&``Anechoic''&Clean&20 sec.&1920 \\ 
Train-target &``Test''&6 to 30dB&20 sec.&1920\\ 
Train-base &``Reverb-C''&6 to 30dB&20 sec.&1920\\ 
Test &``Test''&20dB&20 sec.&300 \\ \midrule
Train-echo-rev&``Echo'' &6 to 30dB&2.5 sec.&1920 \\
Train-echo-anec&``Anechoic''&Clean&2.5 sec&1 \\ 
Test-echo &``Echo''&20dB&2.5 sec.&5 \\ \bottomrule
\end{tabular}
}
\vspace{-13pt}
\end{table}
\vspace{8pt}
\subsection{Experimental setup}
\vspace{-3pt}
{\bf Dataset synthesis:} 
A dataset for training and evaluation was synthesized using IR recordings and dry sounds of the FOA-MEIR dataset. The dataset consists of eight splits as shown in Table~\ref{tb:split}. 
The above five split are constructed in the same way as in the existing SELD dataset dealing with 
stationary polyphonic sound sources~\cite{dataset2019}.
Here, the maximum number of overlapping sources was set to 2, and the clip-wise average of the signal-to-noise ratio (SNR) was randomly set between 6 to 30 dB.
The below three splits in Table~\ref{tb:split} contain sound clips assuming observed echo $\bm{h}_{\mathcal{E}}$.
These observed echoes were synthesized by convolving the 20 ms swept-sine signal~\cite{echo1} with the IR of the ``Echo'' subset of FOA-MEIR.
The ``Train-echo-rev'' is composed of the same number of clips as ``Train-rev''. Each clip is synthesized with the same environment and SNR as ``Train-rev'', but different noise is added. This is because the observed signal $\bm{x}_{\mathcal{E}}$ and the observed echo $\bm{h}_{\mathcal{E}}$ are asynchronous.
The "train-echo-anec" split contains only one clip synthesized using an IR recording in anechoic room, and that clip contains only the direct sound of the swept-sine signal.

\vspace{2pt}
\noindent
{\bf Hyper parameter:} 
 For the short time Fourier transform of $\bm{x}_{\mathcal{E}}$ and $\bm{h}_{\mathcal{E}}$, 2048 and 1024-point Hanning windows with 960 and 512-point shifts were used, respectively.  The dimension of the Mel filter bank was 64. All model parameters in the main branch are the same as in the DCASE2020 baseline model~\cite{dataset2020}. As shown in Fig.~\ref{fig:architecture}, the Echo AE encoder has 4 CNN blocks; the number of CNN filters of each block was (16, 32, 64, 4), the kernel size was 3, the stride and padding were both 1, the max-pooling size was (2, 2) for all blocks. The parameters of the Upsampling CNN-block of The echo AE decoder are symmetric to the encoder. Dimension of $\bm{z}_{\mathcal{E}}$ was $D_{\mbox{\scriptsize echo}}=16$. The number of linear layers of the domain classifier was 3; the input dimension was 512, the hidden dimensions were 512, 128, and the output dimension was 1. In the GRL layer, scale factor $\lambda_{\mbox{\scriptsize grl}}$ was changed during training using the following schedule as in~\cite{grl,doaadap2}:
\vspace{-9pt}
\begin{equation}
\lambda_{\mbox{\scriptsize grl}} =\bar{\lambda}_{\mbox{\scriptsize grl}}\left(\frac{2}{1 + \exp(-\gamma p)}-1\right),
\vspace{-3pt}
\end{equation}
where $p=\frac{\rm epoch}{\rm max epoch}$, $\gamma=10$, and $\bar{\lambda}_{\mbox{\scriptsize grl}}=0.01$. The balance parameter of the loss function $\lambda_{\mbox{\scriptsize doa}},\lambda_{\mbox{\scriptsize seld}},\lambda_{\mbox{\scriptsize domain}}$, and $\lambda_{\mbox{\scriptsize echo}}$ was 100, 3.0, 1.0 and 0.01, respectively. The batch size was 64, half of which was data from an anechoic room and half from reverberant environments. The ADAM optimizer was used for training with an initial learning rate equals 0.01~\cite{Adam}. Training was concluded with 100 epochs.

\vspace{3pt}
\noindent
{\bf Comparison method and evaluation metrics:}
To evaluate the effectiveness of the proposed method, the following six conditions were compared. In the following, the base model refers to the DCASE2020 baseline model that removes the domain classifier and the echo AE from the proposed architecture.

\setlength{\leftmargini}{15pt}    
\begin{description}
\vspace{-2pt}
 \setlength{\parskip}{0cm} 
 \setlength{\itemsep}{0.03cm} 

\item[(A) Target (Oracle):] Training the base model by using the ``Train-target'' split. This is an oracle condition that IR recordings of the unknown environments are available as training data.
\item[(B) Source:] Training the base model by using the ``Train-anec''. This setting is the worst case that any IR recordings in reverberant environments are not available as training data.
\item[(C) Baseline:] Training the base model using the ``train-base'' and ``train-anec''. The ``train-base'' is a conventional dataset that does not employ the proposed CASM scheme, and this result serves as a baseline for our method.
\item[(D) NoAdap:] Training the base model by using ``Train-rev'' and ``Train-anec''.
\item[(E) DAT:] Training the proposed model without echo embedding $\bm{z}_{\mathcal{E}}$ by using ``Train-rev'' and ``Train-anec''.
\item[(F) Proposed:] Training the proposed model by using ``Train-rev'', ``Train-anec'', ``Train-echo-rev'', and ``Train-echo-anec''.
\end{description}

To independently evaluate the effectiveness of the proposed method for DOA estimation and SED, SELD is evaluated with individual metrics for SED and DOA estimation~\cite{Metrics,Metrics2}. For SED, we use the one-second segment-based F-score (F) and error rate (ER) calculated. For DOA estimation, we use the frame-wise metrics of DOA error (DE) that is the average angular error in degrees.
In addition, frame recall (FR) is calculated as the recall of the number of active source estimations. Among these four metrics, higher F and FR and lower ER and DE indicate better performance.

\begin{table}[t!]
\centering
\vspace{-7pt}
\caption{SELD performances. DE, FR, F, and ER denote DOA error, Frame recall, F-value and Error rate of SED, respectively.}
\label{tb:result}
\begin{tabular}{@{}l|cccc@{}}
\toprule
System      & DE$\downarrow$ & FR$\uparrow$ & F$\uparrow$ & ER$\downarrow$ \\ \midrule
(A) Target (Oracle)   &6.1 &95.4 &95.4 &8.8 \\ 
(B) Source &11.9 &68.2 &57.3 &94.3 \\ 
(C) Baseline   &11.9 & 89.7& 84.9&25.3 \\ \midrule
(D) NoAdap   &8.9& 94.5&93.9&11.0\\
(E) DAT   &8.8 & 94.5&94.1 & 10.8\\
(F) Proposed &{\bf 8.4} &{\bf 94.6}& {\bf 94.4}& {\bf 10.5}\\ 
\bottomrule
\end{tabular}
\vspace{-17pt}
\end{table}

\vspace{-10pt}
\subsection{Result}
\vspace{-4pt}
Table~\ref{tb:result} shows the SELD performance of the proposed method and the comparison methods. First, the proposed method (F) outperforms the comparison methods (B) to (E) in all the SELD metrics. Secondly, comparing (C) and (D), even though the number of IR measurements used in (C): 648 is larger than (D): 504, (D) achieves better performance. This fact shows that the proposed data collection scheme, CASM, which combines data from an anechoic room and sparse data from a multi reverberant environment, is more suitable for training a robust SELD model for unknown environments even without DAT and EAR. Comparing (D), (E), and (F), the introduction of DAT and EAR resulted in a stepwise improvement in the performance of DOA estimation, and signs of improvement were also observed for the other metrics. These results indicate that EAR is effective in adapting SELD to unknown environments.

Fig.~\ref{fig:embeddings} shows the t-SNE~\cite{tsne} visualization of the intermediate feature $\bm{f}_{\mathcal{E}^*}$ and $\bm{f}^{'}_{\mathcal{E}^*}$ obtained by (D), (E), and (F) methods when inputting the observed signals of the unknown and anechoic environments. For each method, the left figure shows $\bm{f}_{\mathcal{E}^*}$, and the right figure shows $\bm{f}^{'}_{\mathcal{E^{*}}}$. From the comparison of the distribution of $\bm{f}^{'}_{\mathcal{E}^{*}}$, we can see that the distributions of the anechoic environment, and the unknown environments are different in (D), while the distributions of all environments are mixed in (E) and (F). This suggests that (E) and (F) are successful in suppressing noise and reverberation at the feature level. In addition, when comparing the distributions of $\bm{f}_{\mathcal{E}^{*}}$ in (E) and (F), the distributions in (F) are clearly separated by the environment. It means that the denoising and dereverberation is not performed in $\mathcal{F}$, but is mainly performed in the $\mathcal{R}$. This implies that the EAR prioritizes feature refinement based on the observation of echoes, rather than the knowledge obtained from the training data. This property of the EAR is expected to be good for the robust SELD because feature refinement based on knowledge from limited training data could be useless in unknown environments.

\vspace{-5pt}
\section{Conclusion}
\vspace{-6pt}
In this study, we proposed echo-aware feature refinement (EAR) for the robust SELD system in unknown environments.
EAR associates the spatial cues in an unknown environment obtained through the echo measurement with feature refinement in domain adversarial training manner.
The validation experiments using our FOA-MEIR dataset confirmed that EAR improves the SELD performance in unknown environments.
Therefore, we conclude that the proposed EAR is effective for adaptation of SELD in unknown environment. 
\clearpage

\bibliographystyle{IEEEbib}
\bibliography{refs}

\end{document}